\documentclass[aps,prc,twocolumn,amsfonts,amsmath,amssymb,showpacs,superscriptaddress,floatfix,nofootinbib]{revtex4-1}

\usepackage{graphicx}
\usepackage{longtable}
\usepackage{bm}
\usepackage{multirow}
\usepackage{hyperref}
\usepackage{ulem}
\hypersetup{colorlinks,citecolor=blue,filecolor=blue,linkcolor=blue,urlcolor=blue}
\usepackage{color}
\usepackage{dcolumn}

\begin{document}

\title{Rearrangement of valence neutrons in the neutrinoless double-{\boldmath $\beta$} decay of {\boldmath$^{136}$}Xe}

\author{S.~V.~Szwec}
\affiliation{School of Physics and Astronomy, University of Manchester, Manchester M13 9PL, United Kingdom}
\author{B.~P.~Kay}
\email[E-mail: ]{kay@anl.gov}
\affiliation{Physics Division, Argonne National Laboratory, Argonne, Illinois 60439, USA}
\author{T.~E.~Cocolios}
\altaffiliation[Present address: ]{KU Leuven, Instituut voor Kern- en Stralingsfysica, B-3001 Leuven, Belgium}
\affiliation{School of Physics and Astronomy, University of Manchester, Manchester M13 9PL, United Kingdom}
\author{J.~P.~Entwisle}
\affiliation{School of Physics and Astronomy, University of Manchester, Manchester M13 9PL, United Kingdom}
\author{S.~J.~Freeman}
\affiliation{School of Physics and Astronomy, University of Manchester, Manchester M13 9PL, United Kingdom}
\author{L.~P.~Gaffney}
\affiliation{School of Engineering and Computing, University of the West of Scotland, Paisley PA1 2BE, United Kingdom}
\author{V.~Guimar\~{a}es}
\affiliation{Instituto de F\'{i}sica, Universidade de S\~{a}o Paulo, Caixa Postal 66318, S\~{a}o Paulo 05315-970, S\~{a}o Paulo, Brazil}
\author{F.~Hammache}
\affiliation{Institut de Physique Nucl\'eaire d'Orsay, CNRS-IN2P3, Universit\'e Paris-Sud, Universit\'e Paris-Saclay, 91406 Orsay Cedex, France}
\author{P.~P.~McKee}
\affiliation{School of Engineering and Computing, University of the West of Scotland, Paisley PA1 2BE, United Kingdom}
\author{E.~Parr}
\affiliation{School of Engineering and Computing, University of the West of Scotland, Paisley PA1 2BE, United Kingdom} 
\author{C.~Portail}
\affiliation{Institut de Physique Nucl\'eaire d'Orsay, CNRS-IN2P3, Universit\'e Paris-Sud, Universit\'e Paris-Saclay, 91406 Orsay Cedex, France}
\author{J.~P.~Schiffer}
\affiliation{Physics Division, Argonne National Laboratory, Argonne, Illinois 60439, USA}
\author{N.~de S\'er\'eville}
\affiliation{Institut de Physique Nucl\'eaire d'Orsay, CNRS-IN2P3, Universit\'e Paris-Sud, Universit\'e Paris-Saclay, 91406 Orsay Cedex, France}
\author{D.~K.~Sharp}
\affiliation{School of Physics and Astronomy, University of Manchester, Manchester M13 9PL, United Kingdom}
\author{J.~F.~Smith}
\affiliation{School of Engineering and Computing, University of the West of Scotland, Paisley PA1 2BE, United Kingdom}
\author{I.~Stefan}
\affiliation{Institut de Physique Nucl\'eaire d'Orsay, CNRS-IN2P3, Universit\'e Paris-Sud, Universit\'e Paris-Saclay, 91406 Orsay Cedex, France}

\date{\today}

\begin{abstract}

A quantitative description of the change in ground-state neutron occupancies between $^{136}$Xe and $^{136}$Ba, the initial and final state in the neutrinoless double-$\beta$ decay of $^{136}$Xe, has been extracted from precision measurements of the cross sections of single-neutron adding and -removing reactions. Comparisons are made to recent theoretical calculations of the same properties using various nuclear-structure models. These are the same calculations used to determine the magnitude of the nuclear matrix elements for the process, which at present disagree with each other by factors of  2 or 3. The experimental neutron occupancies show some disagreement with the theoretical calculations.

\end{abstract}


\maketitle

%%%%%%%%%%%%%%%%
{\it Introduction.} An observation of neutrinoless double-$\beta$ ($0\nu2\beta$) decay~\cite{furry} is one of the most tantalizing prospects in contemporary physics; it would inform us that lepton number is not conserved and that  neutrinos are Majorana fermions~\cite{elliott}. In the $0\nu2\beta$-decay process, two neutrons become two protons, thus rearranging the occupancy of protons and neutrons about the orbitals active in the ground states of the parent and daughter of the decay. The rate at which the decay occurs is inversely proportional to the square of the nuclear matrix element, which in turn is proportional to the effective neutrino mass. Currently, the discrepancies between calculations of the nuclear matrix elements using different models are large, around a factor of 2--3 for any given candidate~\cite{elliott}, which corresponds to up to an order of magnitude in the estimated half-life.

While there is no simple experimental probe that connects the same initial and final states seen in $0\nu2\beta$ decay, there are other nuclear-structure properties that can provide important constraints on the calculations used to determine the nuclear matrix elements~\cite{freemanreview}. One such property is the occupancy of valence nucleons in the ground states of the parent and daughter nuclei, and importantly how these change when two neutrons decay into two protons.

Several recent experiments have resulted in descriptions of single-nucleon occupancies for four of the favorable $0\nu2\beta$-decay candidates: \mbox{$^{76}$Ge~$\rightarrow$~$^{76}$Se~\cite{schifferge,kayge}}, $^{100}$Mo~$\rightarrow$~$^{100}$Ru~\cite{freemanmo}, $^{130}$Te~$\rightarrow$~$^{130}$Xe system~\cite{kayxe}, and  $^{136}$Xe~$\rightarrow$~$^{136}$Ba systems~\cite{entwisle}. The current work relates to the last one. Specifically, it builds on Ref.~\cite{entwisle}, which describes the proton occupancies for the $A=136$ system, by providing a quantitative description of the ground-state neutron vacancies and their change between $^{136}$Xe and $^{136}$Ba.

The $^{136}$Xe isotope is the subject of the EXO(-200) and KamLAND-Zen experiments in search of a $0\nu2\beta$ decay signal. These experiments currently set the most stringent limits on the half-life for the $0\nu2\beta$ decay of $^{136}$Xe, being $T^{0\nu}_{1/2}>1.1\times10^{25}$~y~\cite{exo} and $>1.9\times10^{25}$~y~\cite{kamland}, respectively. Just prior to publication of the present work, KamLAND-Zen published results improving on this, placing a limit of  $T^{0\nu}_{1/2}>1.07\times10^{26}$~y~\cite{kamland2}. There are several key properties that make this an attractive candidate: it has the longest $T^{2\nu}_{1/2}$ of all practical candidates at $2\times10^{21}$~y~\cite{exo2v}, a moderately high $Q$ value of 2458~keV~\cite{redshawxe}, and a natural abundance of 8.86\%. The nuclear matrix elements for this candidate currently vary between approximately 1.55 and 3.79 (dimensionless), depending on which nuclear-structure model and other assumptions are used (see, for example, Ref.~\cite{neacsu}). This degree of variation is similar to that seen with other candidates.

To characterize the ground-state neutron occupancies and vacancies of $^{136}$Xe and $^{136}$Ba, overlaps from the neutron-removing and -adding reactions are required. The active orbitals important for these systems are those between $50<N<82$, being $0g_{7/2}$, $1d$, $2s_{1/2}$, and $0h_{11/2}$. These are populated via $\ell=4$, 2, 0, and 5 transfer, respectively. In order to extract reliable spectroscopic factors (reduced cross sections) from the measured cross sections in transfer reactions, it is important to consider momentum matching. In this study, both the ($d$,$p$) and ($\alpha$,$^3$He) reactions were carried out at incident beam energies of a few MeV/u above the Coulomb barrier to probe the neutron vacancy. Similarly, the ($p$,$d$) and ($^3$He,$\alpha$) reactions were used to probe the neutron occupancy. The ($d$,$p$) and ($p$,$d$) reactions are well matched for $\ell=0$ and 2 transfer, while the ($\alpha$,$^3$He) and ($^3$He,$\alpha$) reactions, with larger differences between their incoming and outgoing momenta than the ($d$,$p$) and ($p$,$d$) reactions, are better matched for $\ell=4$ and 5 transfer. This approach has been used in many studies, but most rigorously discussed in Refs.~\cite{schifferni1,schifferni2}. 

There are data in the literature~\cite{nndc} presenting information on single-nucleon overlaps for some of the barium and xenon isotopes. The principal motivation for the additional measurements discussed here lies in the fact that no previous studies used reactions that suitably probe the high-$j$ $0g_{7/2}$ and $0h_{11/2}$ strength, nor is there a consistent dataset on adding and removing reactions such that the Macfarlane and French sum rules~\cite{macfarlane} can be used to inform the analysis. For a given $j^{\pi}$, this simple sum relates the strength seen in nucleon-removing reactions, which probes the occupancy of the orbital, and the strength seen in nucleon-adding reactions, which probes the vacancy of the orbital, to the total degeneracy of the orbital. It provides a consistency check when probing several systems in the same shell-model space, a normalization to allow for comparison to shell-model calculations (see, for example, Ref.~\cite{schifferni1}), and a quantitative description of how the occupancy of orbitals changes from one isotope to another.

An important consideration for the study of the nuclear structure of the $^{136}$Xe~$\rightarrow$~$^{136}$Ba $0\nu2\beta$ decay is whether or not $N=82$ can be considered a `good' closed shell. More explicitly, this question might be couched in terms of the extent to which the neutron $0g_{7/2}$, $1d_{5/2,3/2}$, $2s_{1/2}$, and $0h_{11/2}$ orbitals are full and the neutron $1f_{7/2}$, $2p_{3/2,1/2}$, $0h_{9/2}$, and $0i_{13/2}$ orbitals are empty. There are data that seem to suggest $N=82$ is a good shell closure in this region. From transfer-reaction data this can be answered, to some degree, by seeing whether there is strength attributable to the negative-parity orbitals from above $N=82$ in neutron-removal reactions on $N=82$ isotones, such as ($p$,$d$) and ($d$,$t$). Jolly and Kashy~\cite{jolly}, reporting on the ($p$,$d$) reaction on $^{138}$Ba and $^{140}$Ce, comment that ``we do not see any 7/2$^-$, 3/2$^-$, or 1/2$^-$ states with any measurable intensity.'' This is confirmed in ($p$,$d$)-reaction studies on $^{136}$Xe~\cite{sen} and $^{138}$Ba~\cite{chaumeaux}, which  also conclude that $N=82$ appears to be a good closed shell, and in more recent studies using both the ($p$,$d$) and ($^3$He,$\alpha$) reactions on the stable even-$A$ $N=82$ isotones~\cite{howard}. There are also ($d$,$t$)-reaction studies that support this~\cite{schneid,moore}. 

Similar conclusions can be drawn from neutron-adding reactions. In this case, one looks for whether there are excitations attributable to the positive-parity orbitals from below $N=82$ seen in neutron-adding reactions on $N=82$. Measurements on $^{136}$Xe are limited to the ($d$,$p$) reactions, done twice in normal kinematics~\cite{schneid,moore} and twice in inverse kinematics~\cite{kraus,kayxe136}. The work of Ref.~\cite{kraus} claims there is some 5/2$^+$ strength at 1.87~MeV, though this is not supported in other studies. For $^{138}$Ba, Refs.~\cite{ehrenstein,ipson} see no evidence of positive-parity states (aside from 13/2$^+$, which lies above $N=82$) using the ($d$,$p$) reaction. For the heavier $N=83$ isotones, such as Ce, Nd, and so on, there are some indications of weak fragments of 3/2$^+$ or 5/2$^+$ at several MeV excitation with strengths $\ll1$\% of the total strength~\cite{park}, though assignments are tentative.  

Thus, there is sufficient evidence to support the fact that for $^{136}$Xe and $^{138}$Ba, $N=82$ is a good closed shell and that all orbitals below are fully occupied and those above $N=82$ are empty. In the $0\nu2\beta$ decay of $^{136}$Xe, the rearrangement of the neutrons that occurs in the decay is wholly described by the occupancy of the valence neutrons of $^{136}$Ba. A detailed description of these valence neutron occupancies is not available from the literature.
 
There have been no neutron-removal reactions carried out on $^{136}$Ba. The ($d$,$p$) reaction has been studied previously at an incident energy of 12~MeV in a systematic study across all stable even-$A$ Ba isotopes by von Ehrenstein {\it et al.}~\cite{ehrenstein}. In the present paper, we report on measurements of the ($d$,$p$), ($p$,$d$), ($\alpha$,$^3$He), and ($^3$He,$\alpha$) reactions on $^{136}$Ba, which are used to extract single-neutron occupancies and thus quantify the change in the ground-state valence-nucleon occupancies in the $0\nu2\beta$ decay of $^{136}$Xe. Additional measurements of the same reactions were carried out on $^{134}$Ba to provide important cross checks in the analysis. 

%%%%%%%%%%%%%%%%

{\it Measurement.} Two separate experiments were carried out at the Tandem-Alto facility at the Institut de Physique Nucl\'{e}aire d'Orsay. The first measurements were of the ($\alpha$,$^3$He) reaction at 40.1~MeV and of the ($^3$He,$\alpha$) reaction at 32.0~MeV, both on targets of $^{134,136}$Ba. These exploited the high terminal voltages available with the IPN Orsay tandem. The second measurements, on the same targets, were of the ($d$,$p$) reaction at 15~MeV and the ($p$,$d$) reaction at 23~MeV. 

In all cases, the beams were delivered by the tandem accelerator and incident on the targets in the scattering chamber of an Enge split-pole spectrometer~\cite{spencer}. This was used to momentum analyze the outgoing ions. Their position, energy loss, and residual energy were recorded at the focal plane of the spectrometer using a position-sensitive gas chamber followed by a $\Delta E$ ionization chamber and a plastic scintillator~\cite{markham}. This allowed for the outgoing ions of interest to be isolated from other reaction products and the $Q$-value spectra to be constructed.

\begin{figure}
\centering
\includegraphics[scale=0.7]{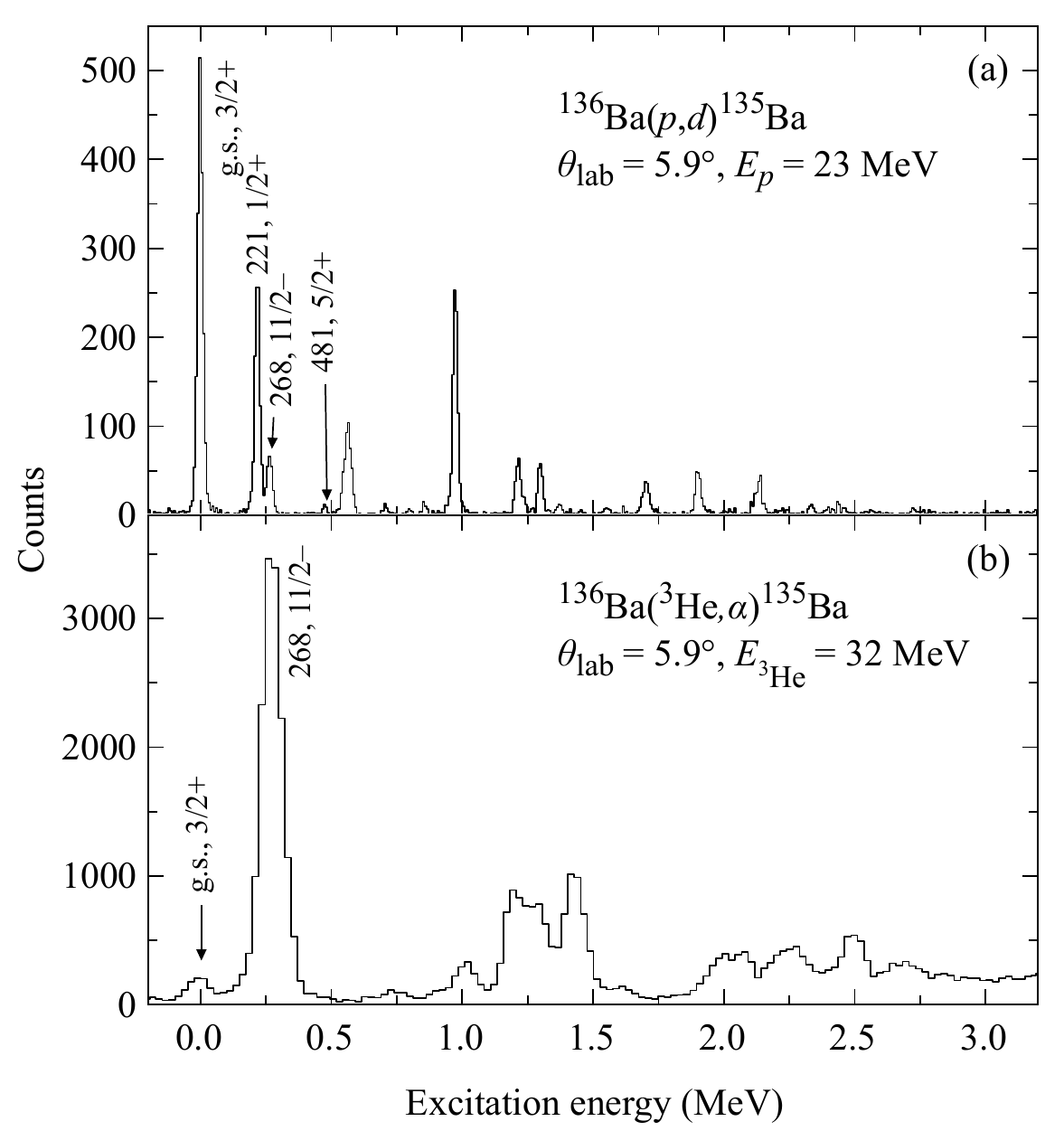}
\caption{\label{fig1} (a) Outgoing deuteron spectrum measured at $\theta_{\rm lab}=5.9^{\circ}$ following the $^{136}$Ba($p$,$d$)$^{135}$Ba reaction at an incident proton energy of 23~MeV and (b) the outgoing $^4$He spectrum measured at the same angle following the $^{136}$Ba($^3$He,$\alpha$)$^{135}$Ba reaction at 32~MeV. Some relevant states are labeled by their energies in keV and spin-parity assignments.}
\end{figure}

The targets were prepared from enriched barium oxides and mounted on a carbon backing. The targets were of thickness $\sim$50-75~$\mu$g/cm$^2$ and the backings $\sim$40~$\mu$g/cm$^2$. To calibrate the product of the target thickness and the spectrometer aperture (nominally 1.6~msr), the $^4$He elastic scattering yield was measured at an incident beam energy of 15~MeV and at a laboratory angle of 20.9$^{\circ}$. The angle was set by markers on the track of the spectrograph to the nearest degree, e.g., 20$^{\circ}$ in this case. An additional 0.9$^{\circ}$ is added due to a well-defined offset of the aperture. In this regime, optical-model calculations show that the elastic scattering cross section is within one percent of Rutherford scattering, and an uncertainty in the angle of 0.1$^{\circ}$ corresponds to a $\sim$1\% uncertainty in cross section. The beam current was measured using a Faraday cup downstream of the target and integrated throughout each run. This, along with the calibration of the target thickness and aperture product, allowed the extraction of absolute cross sections from the transfer-reaction yields. To minimize systematic uncertainties the spectrometer aperture was fixed throughout each experiment, being the same for the elastic-scattering and the transfer-reaction measurements, the same targets were used, and the same settings on the beam-current integrator were used. Example spectra are shown in Fig.~\ref{fig1}, which highlight the effects of momentum matching. The $Q$-value resolution was $\sim$50~keV full width at half-maximum for protons from the ($d$,$p$) reaction, $\sim$40~keV for deuterons from the ($p$,$d$) reaction, $\sim$100~keV for $\alpha$ particles from the ($^3$He,$\alpha$) reaction, and $\sim$70~keV for $^3$He ions from the ($\alpha$,$^3$He) reaction. 

In the first experiment, the neutron-adding ($\alpha$,$^3$He) reaction was carried out at 40.1~MeV at laboratory angles of 5.9$^{\circ}$ for $^{134}$Ba, and 5.9$^{\circ}$ and 10.9$^{\circ}$ for $^{136}$Ba. The forward-most angle of 5.9$^{\circ}$ was a compromise between the high rate in the focal-plane detector and being close to the maximum in the $\ell=4$ and 5 angular distributions---where the spectroscopic factors are most reliably extracted. A second angle was chosen for $^{136}$Ba to reveal peaks that were otherwise obscured by contaminants. This was not necessary in the case of $^{134}$Ba.

The spins and parities of the states of interest in this study are generally well known with robust  assignments. For example, a detailed high-resolution study of the ($\vec{d}$,$p$) reaction on $^{132}$Ba~\cite{suliman} has been carried out. While this does not provide the required information for this work, it does provide confirmation of spin and parity assignments for many of the same states probed in the present study of $^{133}$Ba. Similar is true of ($p$,$d$)-reaction studies on $^{138}$Ba~\cite{jolly} for states in $^{137}$Ba. In both cases, they reinforce the assignments made in other measurements, such as $\beta$ decay. The only case where new assignments may be made in the present work is for the high-$j$ states, which would have been weakly populated in previous studies.

\begin{figure}
\centering
\includegraphics[scale=0.65]{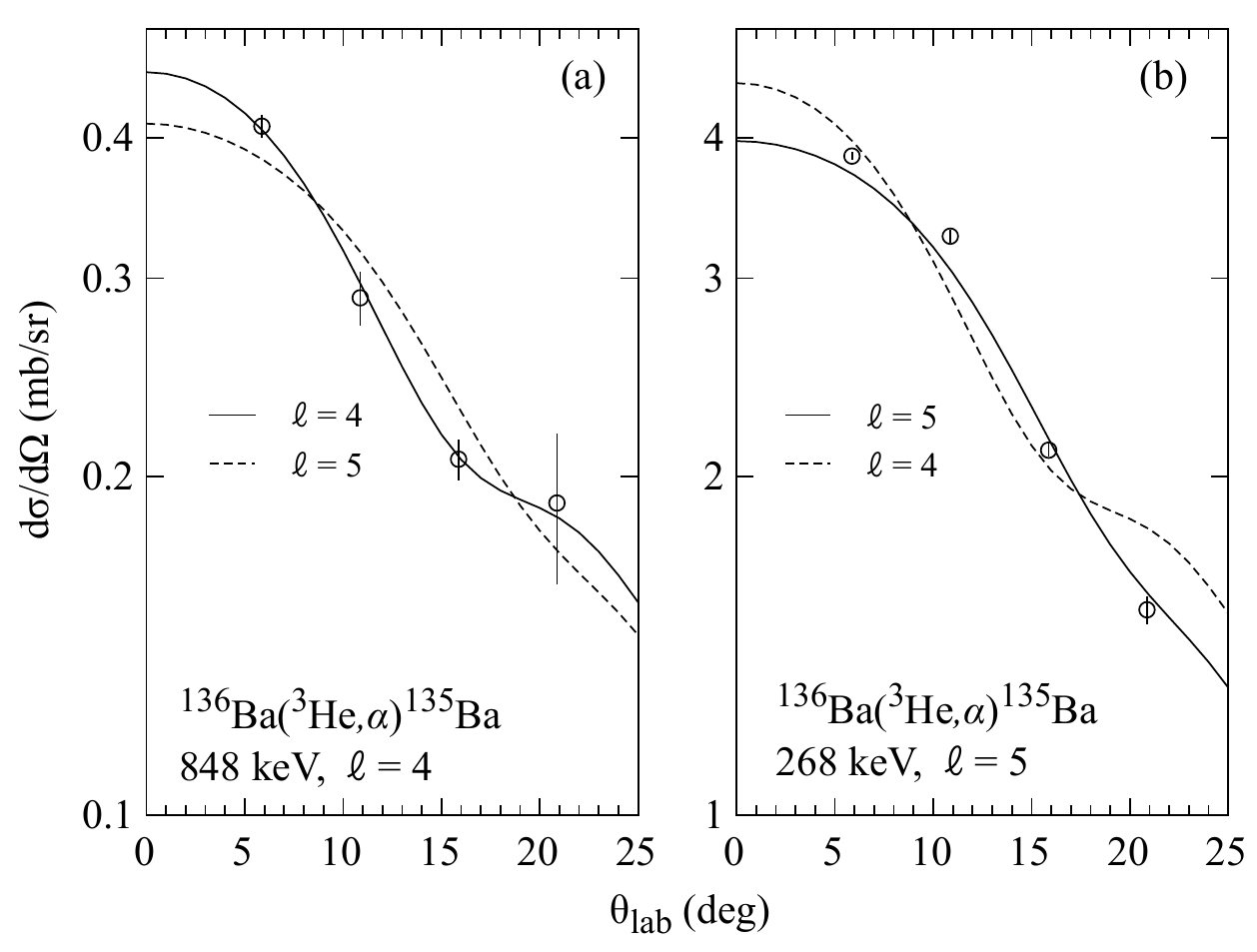}
\caption{\label{fig2} Example angular distributions demonstrating (a) $\ell=4$ and (b) 5 discrimination in the ($^3$He,$\alpha$) reaction on $^{136}$Ba. The solid black lines show DWBA calculations for the assigned $\ell$ value, while the dashed lines show the distribution associated with the alternative $\ell$ value.}
\end{figure}

The neutron-adding reaction probes the vacancy below $N=82$, and the targets are only two and four neutrons short of this. This means that for a given percentage accuracy the adding reactions, determining a smaller quantity, provide a more sensitive measure of the shortfall from $N=82$. Given that the spins and parities are known, we opted to run only at the angle(s) close to the peak of the angular distribution in the neutron-adding measurements. The inverse is true of the neutron-removing ($^3$He,$\alpha$) reaction, whose yields are proportional to the occupancy, which being 28 and 30 neutrons above $N=50$ results in significant yields for this reaction. This measurement was again at angles corresponding to the cross-section maxima for the high-$j$ states, but also at additional angles of 15.9$^{\circ}$ and 20.9$^{\circ}$ to help discriminate between the similar $\ell=4$ and 5 shapes. An example of this is shown in Fig.~\ref{fig2}.

%---------------------------------------------------FIGURE 1------------------------------------------------------
%-----------------------------------------------------------------------------------------------------------------------

In the second experiment, a similar approach was taken with the ($d$,$p$) and ($p$,$d$) reactions. The experimental conditions, such as the spectrometer aperture and beam current integrator, were set to the same values as the previous experiment. The neutron-adding ($d$,$p$) reaction was carried out at $\theta_{\rm lab}=5.9^{\circ}$ and 18.9$^{\circ}$ on both $^{134,136}$Ba. These angles correspond to the maxima of the $\ell=0$ and 2 angular distributions ($\ell=0$ is peaked at 0$^{\circ}$; however, 5.9$^{\circ}$ was the most forward angle at which we could practically run). For the neutron-removing ($p$,$d$) reaction, the same angles were used, again corresponding to maxima in the $\ell=0$ and 2 angular distributions. Mechanical failure of the target prohibited a measurement of the ($p$,$d$) reaction on $^{134}$Ba. During the ($d$,$p$) measurements, a gradual degradation of the targets was noted when the data were analyzed; the counting rate per integrated beam current for a given region of excitation changed as a function of time. The rate of loss of target material for a measurement at a given angle was about 10--20\% per hour. This means that for each target, only the relative yields for the different states at a given angle are meaningful for this part of the experiment and an absolute normalization was not acquired. The implications of this are discussed below.

%%%%%%%%%%%%%%%%

{\it Analysis and results.} In all of the nuclei studied here, the spectra are characterized by a low-lying sequence of 3/2$^+$, 1/2$^+$ and 11/2$^-$ states, which to a large degree define the vacancies in the $\nu1d_{3/2}$, $\nu2s_{1/2}$, and $\nu0h_{11/2}$ orbitals below $N=82$. Both the $^{134,136}$Ba($\alpha$,$^3$He) and ($^3$He,$\alpha$) reactions suggest that the $0h_{11/2}$ strength is carried by one state a few hundred keV from the ground state in all residual nuclei---similar to what was seen in Ref.~\cite{kayxe} for the $^{128,130}$Te isotopes at $N=76$ and 78. The same work saw no evidence for the vacancy of the $\nu0g_{7/2}$ orbital, which would be seen as 7/2$^+$ strength in the neutron-adding ($\alpha$,$^3$He) reaction. Some of the $\nu0g_{7/2}$ strength was seen in the neutron-removing ($^3$He,$\alpha$) reaction over the excitation-energy range studied, but given that this orbital starts filling at $N=50$, some 26--28 neutrons below the isotopes being studied, it is essentially filled and deeply bound. The work of Ref.~\cite{kayxe} reported it would have been sensitive to a vacancy in the $\nu0g_{7/2}$ orbital of about 0.15 nucleons or greater. In the present work, there is some tentative evidence for a small vacancy in the $\nu0g_{7/2}$ orbital seen in weakly populated states at 2.52~MeV excitation energy in $^{136}$Ba($\alpha$,$^3$He).

%---------------------------------------------------FIGURE 2------------------------------------------------------
%-----------------------------------------------------------------------------------------------------------------------

In the data reduction, for each final state the cross section was divided by that calculated with the distorted-wave Born approximation (DWBA) using the code Ptolemy~\cite{ptolemy}. For the scattering potentials, the global optical-model parameters of An and Cai~\cite{an} for deuterons and Koning and Delaroche~\cite{koning} for protons were used for the ($d$,$p$) and ($p$,$d$) reactions. Others~\cite{daehnick,becchettip} were explored and yielded similar results. For both reactions the deuteron wave function was described by the Argonne $\nu_{18}$ potential~\cite{av18}. The target bound-state form factors were taken as a Woods-Saxon plus a spin-orbit derivative term, with $r_0=1.28$~fm, $a=0.65$~fm, $V_{\rm so}=6$~MeV, $r_{\rm so0}=1.1$~fm, and $a_{\rm so}=0.65$~fm. These radial parameters are consistent with those determined in ($e,e'p$) studies~\cite{lapikas}. For the ($\alpha$,$^3$He) and ($^3$He,$\alpha$) reactions, a fixed alpha-particle scattering potential was used with parameters from Ref.~\cite{bassani}, and for the $^3$He ions, global optical-model parameters of Pang {\it et al.}~\cite{pang}. There are fewer available global optical-model parameter sets for alpha particles. Recent values from Ref.~\cite{su} did not reproduce the angular distributions shown in Fig.~\ref{fig2} and so were not used. Several parameter sets of $^3$He ions were explored~\cite{becchettih,trost}, with only small variations between them. Potentials used in analyses of similar studies in the region were also explored, including those used for Sn($\alpha$,$t$)~\cite{schiffersn} and $^{144}$Sm($\alpha$,$^3$He)~\cite{rekstad}, both done with incident beam energies of 40~MeV. For the $^3$He and $^4$He, the projectile wave function was described using the parametrizations of Brida {\it et al.}~\cite{brida} and for the target bound state, the same prescription as for the proton and deuteron-induced reactions.

The experimental cross sections are divided by the DWBA-calculated cross sections to provide the spectroscopic factor, $S_j$. This has to be normalized as the cross sections probed in nucleon-adding and -removing reactions are quenched such that only about 45-65\% of the single-particle strength is seen at low excitation energy and momentum~\cite{lapikas,quench}. This is usually done such that the normalization, $N_j$, for a given orbital is given by
\begin{equation}\label{eqn}
(2j+1)N_j=\Sigma (2j+1)C^2 S_j^{+}+ \Sigma C^2S_j^{-},
\end{equation}
where the spectroscopic factor $S_j^+$ is from the adding reaction,  $S_j^-$ is from the removing reaction, and $C^2$ is the isospin-coupling Clebsch-Gordan coefficient~\cite{isospin}.

%-------------------------------------------------------TABLE 1----------------------------------------------------
\begin{table*}
\caption{\label{tab1} Neutron vacancies from this analysis.}
\newcommand\T{\rule{0pt}{3ex}}
\newcommand \B{\rule[-1.2ex]{0pt}{0pt}}
\begin{ruledtabular}
\begin{tabular}{lccccc}
Isotope\B & $\nu0g_{7/2}$ & $\nu1d$ & $\nu2s_{1/2}$ & $\nu0h_{11/2}$ & Total\footnote{The sums are defined as 4.00 and 2.00 for $^{134,136}$Ba, respectively, and the vacancies for $^{136}$Xe defined as 0.00, as discussed in the text.} \\
\hline
$^{134}$Ba\T&	0.00$^{+0.15}_{-0.00}$	&	1.12$\pm$0.15 &	0.50$\pm$0.15	&	2.38$\pm$0.15	&	4.00	\\

$^{136}$Ba\T\B	&	0.00$^{+0.15}_{-0.00}$	&	0.24$\pm$0.05	&	0.08$\pm$0.02	&	1.68$\pm$0.13	&	2.00	\\

$^{136}$Xe	&	0.00	&	0.00	&	0.00	&	0.00	&	0.00	\\

\hline
$^{136}$Ba$-^{136}$Xe\T	&	0.00$^{+0.15}_{-0.00}$	&	0.24$\pm$0.05	&	0.08$\pm$0.02	& 1.68$\pm$0.13	&	2.00	\\
\end{tabular}
\end{ruledtabular}
\end{table*}
%-----------------------------------------------------------------------------------------------------------------------

%---------------------------------------------------FIGURE 3------------------------------------------------------
\begin{figure*}
\centering
\includegraphics[scale=0.85]{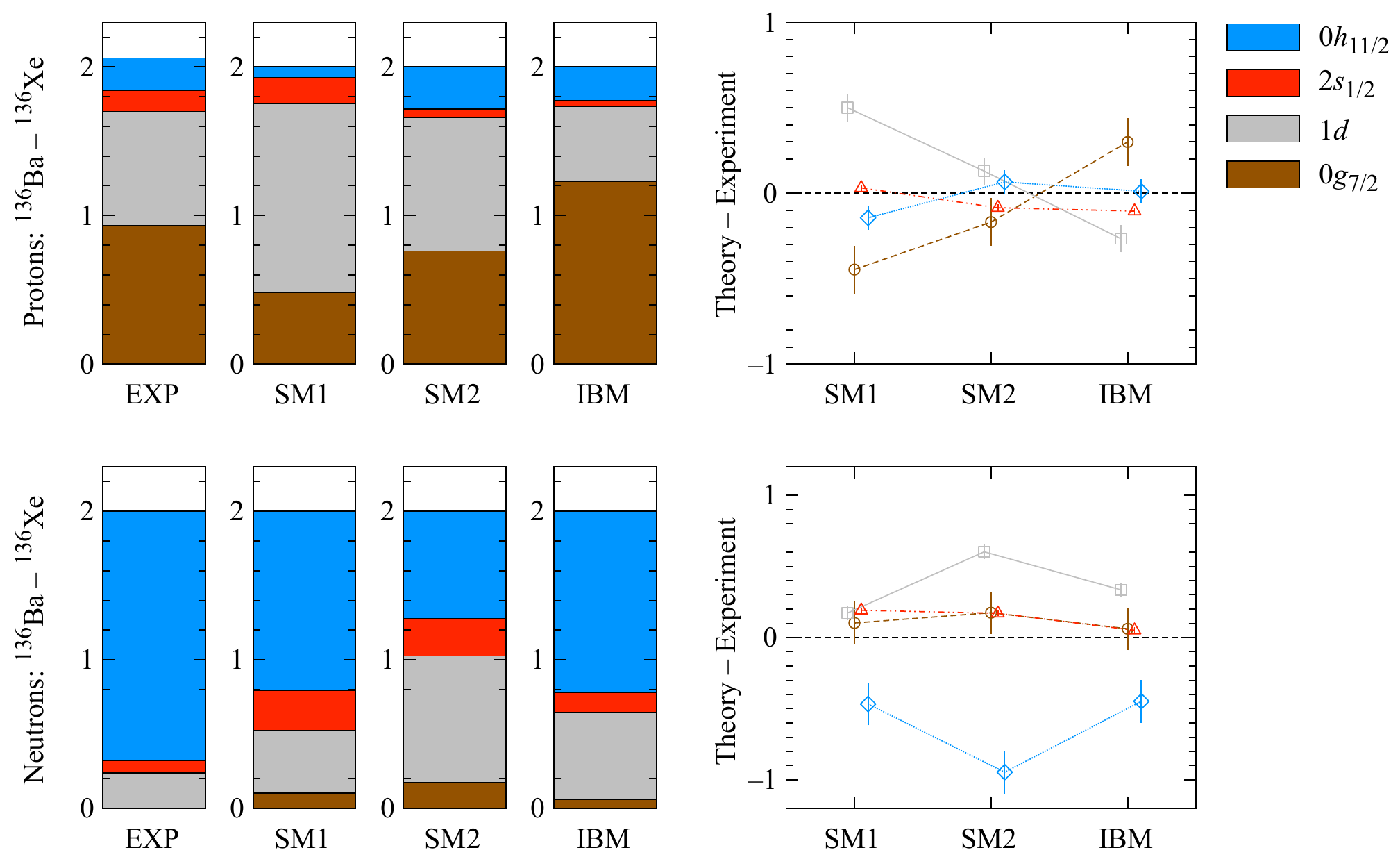}
\caption{\label{fig3}(color online). The change in proton occupancies (top) and neutron vacancies (bottom) in the $0\nu2\beta$ decay of $^{136}$Xe in the bar charts. The proton data are from Ref.~\cite{entwisle} and the neutron data from the present work. The three different theoretical calculations are from the shell model, SM1~\cite{neacsu} and SM2~\cite{menendez}, and the interacting boson model, IBM~\cite{kotila}. The discrepancy between the theoretical calculations and the experimental data is shown on the right where the error bars show the experimental uncertainty.}
\end{figure*}
%-----------------------------------------------------------------------------------------------------------------------

The loss of absolute cross sections for the ($p$,$d$) and ($d$,$p$) data complicates our usual procedures. Equation~\ref{eqn} is used to determine the reaction normalization in the helium-induced reactions for $\nu0h_{11/2}$ strength. This is then used to determine the vacancy of the $\nu0h_{11/2}$ orbital using the data from the ($\alpha$,$^3$He) reaction, which accounts for the majority of the neutron vacancies in $^{134,134}$Ba. The remaining neutron vacancy is shared between the $\nu2s_{1/2}$ and $\nu1d$ orbitals (mainly the $\nu1d_{3/2}$), and to a much lesser extent the $\nu0g_{7/2}$ and $\nu1d_{3/2}$ orbitals, as discussed in more detail below.  This remaining vacancy is attributed to $\ell=0$ and 2 orbitals and the data from the ($d$,$p$) reaction are used to determine the ratio of these two.

Taking the He-induced reactions on $^{136}$Ba for the $j^{\pi}=11/2^-$ as an example, $(2j+1)C^2S^+=0.94$ and \mbox{$C^2S^-=5.99$}. For the adding reaction $C^2$ is 1. For the neutron-removing reaction, a correction should be made to account for the isobaric analog state (IAS), which was not observed in this reaction at the excitation energies we probed. The correction is small in this case, being $1/(2T+1)$, or $1/25$ times the spectroscopic factor for proton removal from the $\pi0h_{11/2}$ orbital. Were the proton $0h_{11/2}$ orbital fully occupied, this would be 12/25, or 0.48 nucleons. Recent work by Entwisle {\it et al.}~\cite{entwisle} measured the $\pi0h_{11/2}$ occupancy to be small, 0.62, about 5\% of the full shell, as would be expected below the $Z=64$ sub-shell closure. The correction to $C^2S^-$ is thus $\sim0.025$, negligible compared to 5.99, and less than the uncertainties. It is, however, included. This then makes $N_j=(6.02+0.94)/12=0.58$. This value of normalization is consistent with other works~\cite{quench}. Applying this yields a vacancy (occupancy) for the $\nu0h_{11/2}$  orbital of 1.68 (10.32). The remaining $2-1.68 = 0.32$ nucleons must be $\nu2s_{1/2}$ and $\nu1d$ (probably $1d_{3/2}$) strength (see comments on the $\nu0g_{7/2}$ below). The ratio between the $1d$ and $2s$ components was taken from the $\theta_{\rm lab}=5.9^{\circ}$ ($d$,$p$) data---this angle chosen over the $10.9^{\circ}$ data to avoid the sharp minimum in the $\ell=0$ angular distribution, which is poorly described by DWBA. This procedure was also carried out for $^{134}$Ba, and with numerous different parametrizations in the DWBA calculations. The rms spread with different parametrizations is included in the estimate of the uncertainties.

The $\nu0g_{7/2}$ vacancy is very small. We set an upper limit in this work of 0.15 nucleons, which is commensurate with both the sensitivity to this strength in our previous work~\cite{kayxe} and to the uncertainties in the present work. A 7/2$^+$ state at 2.53~MeV in $^{137}$Ba has been reported in the ($p$,$d$) study of Ref.~\cite{jolly}, however, it was not seen in the ($p$,$d$)-reaction studies of Refs.~\cite{chaumeaux,howard}, and was surmised to be a weak $\ell=2$ state which would account for its relatively weak population in the ($^3$He,$\alpha$)-reaction study of Ref.~\cite{howard}. In the present work, it is weakly populated at a level of $<$10~$\mu$b/sr, an order of magnitude weaker than the $\nu0h_{11/2}$ strength. An $\ell=4$ transition with this $Q$ value is poorly matched in angular momentum, making the extraction of a spectroscopic factor less reliable. In $^{135}$Ba, a state previously reported to be a tentative 5/2 or 7/2$^+$ assignment at 1.56~MeV was populated in this work with a cross section of $\sim$20~$\mu$b/sr in the ($\alpha$,$^3$He) reaction, again an order of magnitude weaker than the $\nu0h_{11/2}$ strength. From the angular distribution of the ($^3$He,$\alpha$) cross sections, an $\ell=4$ assignment looks probable. Because of the very low cross sections and poor momentum matching, we only assign an upper limit to the vacancy of the $\nu0g_{7/2}$ orbit.

The vacancies for both $^{134}$Ba and $^{136}$Ba are given in Table~\ref{tab1}. For $^{136}$Ba the change in neutron vacancy with respect to $^{136}$Xe is shown in Fig.~\ref{fig3} along with recent data on the proton occupancies~\cite{entwisle}. The cross sections, in absolute units for the ($\alpha$,$^3$He) and ($^3$He,$\alpha$) reactions and in arbitrary units for the ($p$,$d$) and ($d$,$p$) reactions, and spectra for each reaction, are provided in the Supplemental Material~\cite{supmat}. The uncertainties in this work hinge on those associated with the extraction of the $0h_{11/2}$ strength. The systematic uncertainties on the absolute cross sections for the ($\alpha$,$^3$He) and ($^3$He,$\alpha$) are most likely dominated by uncertainties in the angle of the spectrometer, the uniformity of the targets, and the implementation of the Faraday cup and beam current integrator. These are not trivial to estimate, and so we place a conservative estimate of $\sim$20\%. The relative uncertainties on the cross sections, target-to-target, are smaller because all variables were kept the same except for the targets themselves. These were estimated to be around 5\%. Statistical uncertainties on the large peaks, the 11/2$^-$, were $\sim$1\% for the neutron removing and $\sim$5\% for the neutron adding, becoming $>$10\% for peaks with cross sections $\lesssim$20~$\mu$b/sr. For the ($d$,$p$) and ($p$,$d$) reactions, the statistical uncertainties were less than 5\% for peaks with cross sections $>$0.2--4~mb/sr. The uncertainties in the summed strength is driven largely by the DWBA analysis, which yielded a spread of around $\pm$0.15 nucleons in the $\nu0h_{11/2}$ strength. The rms spread in the $\ell=0$ and 2 vacancies based on the DWBA analysis is $\ll0.1$ nucleons and is thus dominated by the uncertainties in the $\nu0h_{11/2}$ derived from the ($\alpha$,$^3$He) and ($^3$He,$\alpha$) data. Because of the relatively small values of the $\nu2s_{1/2}$ and $\nu1d$ strength, we adopt the same uncertainty as the $\nu0h_{11/2}$ orbitals for vacancies $>$0.5 nucleons, and assume 20\% for values $<$0.5 nucleons. The $\nu0g_{7/2}$ vacancy is left as an upper limit and is not included in the sums.

%%%%%%%%%%%%%%%%
{\it Discussion.} With $^{136}$Ba lying just two neutrons away from $N=82$, it is not surprising that the $\nu0h_{11/2}$ accounts for a large fraction of the vacancy as shown in Fig.~\ref{fig3}. Two of the three calculations also show that the $\nu0h_{11/2}$ accounts for most of the vacancy. It is, however, underestimated in all cases, most notably in the latest shell-model calculations (SM2)~\cite{neacsu}. To a small extent, this is offset by the presence of some vacancy in the $0g_{7/2}$ orbital in all calculations, for which we set an upper limit from the experiment comparable to that in all the theoretical calculations. This is similar to what was observed in the case of $^{130}$Te~$\rightarrow$~$^{130}$Xe. All available theoretical calculations show that significant changes in the vacancy of both the $\nu0h_{11/2}$ and $\nu1d$ orbitals occur in the $0\nu2\beta$ decay of $^{136}$Xe, which is in contrast to the experimental data. The changes in the $\nu0g_{7/2}$ and $\nu2s_{1/2}$ orbitals are in relatively good agreement. The change in proton occupancies for both $A=130$ and 136 systems are quite similar~\cite{entwisle}, perhaps owing to the fact that they are both close to the closed $Z=50$ shell. For neutrons, they are quite different. For $A=130$, there is a large {\it change} in the $\nu1d$ occupancy of around 1.25 nucleons, but far less for $A=136$ of around 0.25 nucleons, where the bulk of the change is in the $\nu0h_{11/2}$ orbital. This may simply reflect the proximity of $^{136}$Ba to $N=82$.

%%%%%%%%%%%%%%%
{\it Conclusion.} These results, along with recent results on the proton occupancies from Ref.~\cite{entwisle}, complete a description of the ground-state proton and neutron occupancies for both $^{136}$Xe and $^{136}$Ba, the parent and daughter of the $0\nu2\beta$-decay candidate. Further, it completes the work on three of the most promising candidates, $^{76}$Ge, $^{130}$Te, and $^{136}$Xe. Common to all these is that, in general, there are significant discrepancies between a theoretical description of the ground-state occupancies and how they change in the decay, and the experimental data. This highlights an important deficiency in the calculations. Even if the calculations of the nuclear matrix elements were insensitive to the occupancies, if they do not correctly describe the initial and final states, of which the occupancies provide a description, it is arguable that it is difficult to draw conclusions about the reliability of the nuclear matrix element. The experimental data from this work provide useful comparisons for new theoretical calculations.

%%%%%%%%%%%%%%%%
{\it Acknowledgements.} This measurement (Experiment NI-S-73) was performed at the Tandem ALTO facility at IPN Orsay. The authors wish to thank the operating staff, and the outside participants wish to thank the local staff and administration for their hospitality and assistance. We are indebted to John Greene for preparing targets for these experiments. This material is based upon work supported by the U.S. Department of Energy, Office of Science, Office of Nuclear Physics, under Contract Number DE-AC02-06CH11357, and by the UK Science and Technology Facilities Council.

%-----------------------------------------------------------------------------------------------------------------------


\begin{references}
%-----------------------------------------------------------------------------------------------------------------------

\bibitem{furry} W.~H.~Furry, \href{http://dx.doi.org/10.1103/PhysRev.56.1184}{Phys. Rev. {\bf 56}, 1184 (1939)}.
\bibitem{elliott} S.~R.~Elliott and M.~Franz, \href{http://dx.doi.org/10.1103/RevModPhys.87.137}{Rev. Mod. Phys. {\bf 87}, 137 (2015)} and references within.
\bibitem{freemanreview} S.~J.~Freeman and J.~P.~Schiffer, \href{http://dx.doi.org/10.1088/0954-3899/39/12/124004}{J. Phys. G: Nucl. Part. Phys. 39, 124004 (2012).}
\bibitem{schifferge} J.~P.~Schiffer {\it et al.,} \href{http://dx.doi.org/10.1103/PhysRevLett.100.112501}{Phys. Rev. Lett. {\bf 100}, 112501 (2008)}.
\bibitem{kayge} B.~P.~Kay {\it et al.,} \href{http://dx.doi.org/10.1103/PhysRevC.79.021301}{Phys. Rev. C {\bf 79}, 021301(R) (2009)}. 
\bibitem{freemanmo} S.~J.~Freeman {\it et al.} (unpublished). 
\bibitem{kayxe} B.~P.~Kay {\it et al.,} \href{http://dx.doi.org/10.1103/PhysRevC.87.011302}{Phys. Rev. C {\bf 87}, 011302(R) (2013)}.
\bibitem{entwisle} J.~P.~Entwisle {\it et al.,} \href{http://dx.doi.org/10.1103/PhysRevC.93.064312}{Phys. Rev. C {\bf 93}, 064312 (2016)}.
\bibitem{exo} J.~B.~Albert {\it et al.,} (The EXO-200 Collaboration), \href{http://dx.doi.org/10.1038/nature13432}{Nature {\bf 510}, 229 (2014)}.
\bibitem{kamland} A.~Gando {\it et al.,} (The KamLAND-Zen Collaboration), \href{http://dx.doi.org/10.1103/PhysRevLett.110.062502}{Phys. Rev. Lett. {\bf 110}, 062502 (2013)}.
\bibitem{kamland2} A.~Gando {\it et al.,} (The KamLAND-Zen Collaboration), \href{http:dx.doi.org/10.1103/PhysRevLett.117.082503}{Phys. Rev. Lett. {\bf 117}, 082503 (2016).}

\bibitem{exo2v} N.~Ackerman {\it et al.,} (The EXO Collaboration), \href{http://dx.doi.org/10.1103/PhysRevLett.107.212501}{Phys. Rev. Lett. {\bf 107}, 212501 (2011)}.
\bibitem{redshawxe} M.~Redshaw, E.~Wingfield, J.~McDaniel, E.~G.~Myers, \href{http://dx.doi.org/10.1103/PhysRevLett.98.053003}{Phys. Rev. Lett. {\bf 98}, 053003 (2007)}.
\bibitem{neacsu} A.~Neacsu and M.~Horoi (private communication); and \href{http:/dx.doi.org/10.1103/PhysRevC.91.024309}{Phys. Rev. C {\bf 91}, 024309 (2015)}.
\bibitem{schifferni1} J.~P.~Schiffer {\it et al.,} \href{http://dx.doi.org/10.1103/PhysRevLett.108.022501}{Phys. Rev. Lett. {\bf 108}, 022501 (2012)}.
\bibitem{schifferni2} J.~P.~Schiffer {\it et al.,} \href{http://dx.doi.org/10.1103/PhysRevC.87.034306}{Phys. Rev. C {\bf 87}, 034306 (2013)}.
\bibitem{nndc} \href{http://www.nndc.bnl.gov/ensdf/}{http://www.nndc.bnl.gov/ensdf/}.
\bibitem{macfarlane} M.~H.~Macfarlane and J.~B.~French, \href{http://dx.doi.org10.1103/RevModPhys.32.567}{Rev. Mod. Phys. {\bf 32}, 567 (1960).}
\bibitem{jolly} R.~K.~Jolly and E.~Kashy, \href{http://dx.doi.org/10.1103/PhysRevC.4.1398}{Phys. Rev. C {\bf 4}, 1398 (1971)}.
\bibitem{sen} S.~Sen, P.~J.~Riley, and T.~Udagawa, \href{http://dx.doi.org/10.1103/PhysRevC.6.2201}{Phys. Rev. C {\bf 6}, 2201 (1972)}.
\bibitem{chaumeaux} A.~Chaumeaux, G.~Bruge, H.~Faraggi, and J.~Picard, \href{http://dx.doi.org/10.1016/0375-9474(71)90849-9}{Nucl. Phys. A {\bf 164}, 176 (1971)}.
\bibitem{howard} A.~M.~Howard, Ph.D. thesis, University of Manchester, 2011.
\bibitem{schneid} E.~J.~Scneid and B.~Rosner, \href{http://dx.doi.org/10.1103/PhysRev.148.1241}{Phys. Rev. {\bf 148}, 1241 (1966)}.
\bibitem{moore} P.~A.~Moore, P.~J.~Riley, C.~M.~Jones, M.~D.~Mancusi, and J.~L.~Foster, Jr., \href{http://dx.doi.org/10.1103/PhysRev.175.1516}{Phys. Rev. {\bf 175}, 1516 (1968)}.
\bibitem{kraus} G.~Kraus {\it et al.,} \href{http://dx.doi.org/10.1007/BF01294683}{Z. Phys. {\bf A340}, 339 (1991)}.
\bibitem{kayxe136} B.~P.~Kay {\it et al.,} \href{http://dx.doi.org/10.1103/PhysRevC.84.024325}{Phys. Rev. C {\bf 84}, 024325 (2011)}.
\bibitem{ehrenstein} D.~von~Ehrenstein, G.~C.~Morrison, J.~A.~Nolen, Jr., and N.~Williams, \href{http://dx.doi.org/10.1103/PhysRevC.1.2066}{Phys. Rev. C {\bf 1}, 2066 (1970)}.
\bibitem{ipson} S.~S.~Ipson, W.~Booth, and J.~G.~B.~Haigh, \href{http:/dx.doi.org//10.1016/0375-9474(73)90610-6	}{Nucl. Phys. {\bf A206}, 114 (1973)}.
\bibitem{park} J.~E.~Park, W.~W.~Daehnick, and M.~J.~Park, \href{http://dx.doi.org/10.1103/PhysRevC.15.587}{Phys. Rev. C {\bf 15}, 587 (1977)}.
\bibitem{spencer} J.~E.~Spencer and H.~A.~Enge, \href{http://dx.doi.org/10.1016/0029-554X(67)90684-2}{Nucl. Instrum. Methods {\bf 49}, 181 (1967)}.
\bibitem{markham} R.~G.~Markham and R.~G.~H.~Robertson, \href{http://dx.doi.org/10.1016/0029-554X(75)90122-6}{Nucl. Instrum. Methods {\bf 129}, 131 (1975)}.
\bibitem{suliman} G.~Suliman {\it et al.,} \href{http://dx.doi.org/10.1140/epja/i2009-10847-9}{Eur. Phys. J. A {\bf 41}, 299 (2009)}.
\bibitem{ptolemy} M.~H.~Macfarlane and S.~C.~Pieper, Argonne National Laboratory Report No. ANL-76-11 Rev. 1, 1978 (unpublished).

\bibitem{an} H.~An and C.~Cai, \href{http://dx.doi.org/10.1103/PhysRevC.73.054605}{Phys. Rev. C {\bf 73}, 054605 (2006)}. 
\bibitem{koning} A.~J.~Koning and J.~P.~Delaroche, \href{http://dx.doi.org/10.1016/S0375-9474(02)01321-0}{Nucl. Phys. A {\bf 713}, 231 (2003)}.

\bibitem{daehnick} W.~W.~Daehnick, J.~D.~Childs, and Z.~Vrcelj, \href{http://dx.doi.org/10.1103/PhysRevC.21.2253}{Phys. Rev. C {\bf 21}, 2253 (1980)}.
\bibitem{becchettip} F.~D.~Becchetti, Jr. and G.~W.~Greenlees, \href{http://dx.doi.org/10.1103/PhysRev.182.1190}{Phys. Rev. {\bf 182}, 1190 (1969)}.
\bibitem{av18} R.~B.~Wiringa, V.~G.~J.~Stoks, and R.~Schiavilla, \href{http://dx.doi.org/10.1103/PhysRevC.51.38}{Phys. Rev. C {\bf 51}, 38 (1995).}
\bibitem{lapikas} G.~J.~Kramer, H.~P.~Blok, and L.~Lapik\'as, \href{http://dx.doi.org/10.1016/S0375-9474(00)00379-1}{Nucl. Phys. A {\bf 679}, 267 (2001).}
\bibitem{bassani} G.~Bassani and J.~Picard, \href{http://dx.doi.org/10.1016/0375-9474(69)90601-0}{Nucl. Phys. A {\bf 131}, 653 (1969)}.
\bibitem{pang} D.~Y.~Pang, P.~Roussel-Chomaz, H.~Savajols, R.~L.~Varner, and R.~Wolski \href{http://dx.doi.org/10.1103/PhysRevC.79.024615}{Phys. Rev. C {\bf 79}, 024615 (2009)}.
\bibitem{su} X.-W.~Su and Y.-L.~Han, \href{http://dx.doi.org/10.1103/PhysRev.182.1190}{Int. J. Mod. Phys. E {\bf 24}, 1550092 (2015)}.

\bibitem{becchettih} F.~D.~Becchetti, Jr. and G.~W.~Greenlees in {\it Polarization Phenomena in Nuclear Reactions}, edited by H.~H.~Barschall and W.~Haeberli (University of Wisconsin Press, Madison, WI, 1971), p. 682.
\bibitem{trost} H.-J.~Trost, P.~Lezoch, and U.~Strohbusch, \href{http://dx.doi.org/10.1016/0375-9474(87)90551-3}{Nucl. Phys. A {\bf 462}, 333 (1987)}.

\bibitem{schiffersn} J.~P.~Schiffer {\it et al.,} \href{http://dx.doi.org/10.1103/PhysRevLett.92.162501}{Phys. Rev. Lett. {\bf 92}, 162501 (2004)}.
\bibitem{rekstad} J.~Rekstad, I.~Espe, G.~L\o vh\o iden, J.~R.~Lien, J.~C.~Waddington, C.~Gaarde, J.~S.~Larsen, S.~Van Der Werf, \href{http://dx.doi.org/10.1016/0375-9474(81)90030-0}{Nucl. Phys. A {\bf 369}, 453 (1981)}.

\bibitem{brida} I.~Brida, S.~C.~Pieper, and R.~B.~Wiringa, \href{http://dx.doi.org/10.1103/PhysRevC.84.024319}{Phys. Rev. C {\bf 84}, 024319 (2011)}.
\bibitem{quench} B.~P.~Kay, J.~P.~Schiffer, and S.~J.~Freeman, \href{http://dx.doi.org/10.1103/PhysRevLett.111.042502}{Phys. Rev. Lett. {\bf 111}, 042502 (2013)}.

\bibitem{isospin} J.~P.~Schiffer, $\it {Isospin~in~Nuclear~Physics}$, p. 665 D.~H.~Wilkinson,~editor, North-Holland Publishing Co. 1969.

\bibitem{menendez} J. Men\'endez {\it et al.,} private communication and \href{http://dx.doi.org/10.1016/j.nuclphysa.2008.12.005}{Nucl. Phys. A {\bf 818}, 139 (2009)}.

\bibitem{kotila} J.~Kotila and J.~Barea, \href{https://doi.org/10.1103/PhysRevC.94.034320}{Phys. Rev. C {\bf 94}, 034320 (2016)}. 

\bibitem{supmat} See Supplemental Material, which can be found online at \href{https://doi.org/10.1103/PhysRevC.94.054314}{https://doi.org/10.1103/PhysRevC.94.054314}.


\end{references}
\end{document}